\documentclass[]{aa}
\usepackage{float}
\usepackage{graphicx}
\usepackage{txfonts}
\usepackage{hyperref}
\hypersetup{colorlinks=true, allcolors=blue}
\usepackage{siunitx}
\usepackage{natbib}
\usepackage{rotating}
\usepackage{booktabs,tabularx,ragged2e,makecell}
\usepackage{amssymb}
\usepackage[T1]{fontenc}
\DeclareSIUnit\year{yr}
\DeclareSIUnit\parsec{pc}
\newcolumntype{Y}{>{\RaggedRight\arraybackslash}X}
\makeatletter
\renewcommand*\aa@pageof{, page \thepage{} of \pageref*{LastPage}}
\makeatother
\begin{document} 
\title{Accelerated gas flow along Ophiuchus B44 filament}
\subtitle{Breaking Position-Position-Velocity degeneracy}
\authorrunning{Alves et al.}
\author{J.  Alves\inst{1}, C. Zucker\inst{2}, C. Lada\inst{2}, M. Lombardi\inst{3},  M. Piecka\inst{1}, S. Hutschenreuter\inst{1}, S. Meingast\inst{1}, L. Posch\inst{1}, A. Hacar\inst{1}, K. Tachihara\inst{4}, R. Yamada\inst{5}, C. Swiggum\inst{2},  A. Goodman\inst{2}, R. Wunsch\inst{6}, J. Gro\ss schedl\inst{6}, A. Burkert\inst{7}, F. Heitsch\inst{8}, \\ T. Enßlin\inst{9,10,11,12}}
\mail{joao.alves@univie.ac.at} \institute{University  of Vienna, Department of Astrophysics, T\"urkenschanzstrasse 17, 1180 Vienna, Austria 
  \and Harvard-Smithsonian Center for Astrophysics, Mail Stop 72, 60 Garden Street, Cambridge, MA 02138, USA 
  \and University of Milan, Department of Physics, via Celoria 16, I-20133, Milan, Italy 
  \and Department of Physics, Graduate School of Science, Nagoya University, Furo-cho, Chikusa-ku, Nagoya 464-8601, Japan
  \and Nobeyama Radio Observatory, National Astronomical Observatory of Japan (NAOJ), National Institutes of Natural Sciences (NINS), 462-2 Nobeyama, Minamimaki, Minamisaku, Nagano 384-1305, Japan
  \and Astronomical institute of the Czech Academy of Sciences, Bo\v{c}n\'{i} II 1401, CZ-14100 Prague, Czech Republic
  \and University Observatory Munich (USM) Scheinerstrasse 1, D-81679 Munich, Germany
  \and University of North Carolina at Chapel Hill, Department of Physics and Astronomy, Chapel Hill, NC 27599, USA
  \and Max Planck Institute for Astrophysics, Karl-Schwarzschild-Str. 1, 85748 Garching, Germany
  \and Deutsches Zentrum für Astrophysik, Postplatz 1, 02826 Görlitz, Germany
  \and Ludwig-Maximilians-Universit\"at M\"unchen,
Geschwister-Scholl-Platz 1, 80539 Munich, Germany
  \and Excellence Cluster ORIGINS, Boltzmannstr. 2, 85748 Garching, Germany
  } 
\date{Received 28 June 2026; accepted 26 July 2026}
  \abstract
    {
    Stellar feedback from massive stars in the Upper-Sco has been proposed to have reshaped the gas in the nearby Ophiuchus complex. In this framework, feedback organizes the gas into two filament types based on their orientation relative to the source of feedback: radial (R-type) filaments, aligned radially to the massive stars, and tangential (T-type) filaments, which are orthogonal to the feedback direction. A key prediction of this scenario is that gas within R-type filaments should flow longitudinally away from the massive stars. In this paper, we test this scenario by measuring the three-dimensional gas flow inside the potential R-type filament B44, combining the 3D orientation of the filament from Gaia-based 3D dust maps with radial velocities from $^{12}$CO and $^{13}$CO observations, effectively breaking the Position-Position-Velocity degeneracy. We find that gas flows longitudinally along the B44 filament away from the massive stars in Upper-Sco with both tracers yielding consistent velocity fields. This result identifies B44 as an R-type filament formed by stellar feedback from Sco-Cen with an implied filament assembly timescale of $\sim$3~Myr, well within the age of the Upper-Sco massive stars. Moreover, we find that the gas motion along B44 and away from the massive stars is accelerated with $a\sim$1.8~km/s/Myr ($\sim 6 \times 10^{-11}$~m/s$^2$). This acceleration is compatible with the accelerations recorded along the Sco-Cen cluster chains over the past $\sim$15~Myr, indicating that B44 is likely a present-day, gas-phase counterpart of the same feedback-driven process that produced those stellar sequences.  We further find evidence for a shock at the wind-facing head of the filament, with a deprojected flow Mach number of $\sim$2 and a matching density jump, and we discuss a possible role for the magnetic field in confining the tail. Our findings demonstrate that Gaia 3D dust maps can lift the line-of-sight ambiguity intrinsic to PPV spectral data, enabling direct deprojection of the gas velocity field in coherent filaments.
    }

   \maketitle

\begin{figure*}
    \centering
    \includegraphics[width=\linewidth]{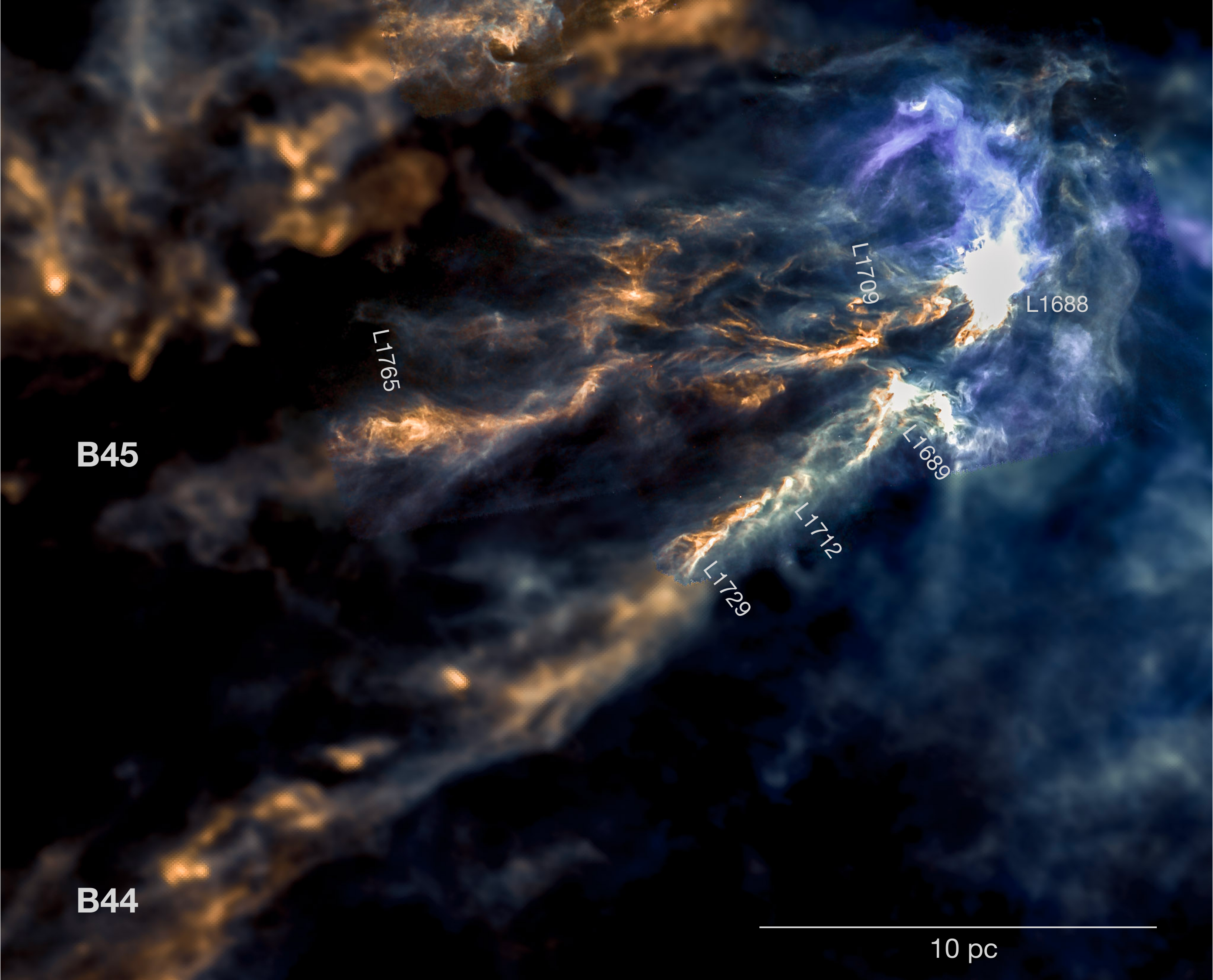}
    \caption{Three-color composite of the B44 filament and its surroundings from ESA \textit{Herschel} observations (blue: 250~$\mu$m, green: 350~$\mu$m, red: 500~$\mu$m) embedded in the lower resolution ESA \textit{Planck} observations (blue: extrapolated 250~$\mu$m, green: 350~$\mu$m, red: 500~$\mu$m). The color coding roughly traces dust temperature: blue-white regions indicate warmer dust ($\sim30$~K) associated with the star-forming L1688 and head of the filament L1689 and the influence of the nearby Upper-Sco OB association, while orange-red tones trace colder dust ($\sim10$ K) in the body and tail of B44. The elongated head--tail morphology of the filament, stretching from the dense head at upper right toward the lower left, is the structure predicted for R-type filaments shaped by stellar feedback \citep{alves_hp2_2025}. The diffuse blue emission pervading the region traces warm dust heated by the radiation field of the Upper-Sco massive stars.}
    \label{fig:herschel}
\end{figure*}

\section{Introduction}
Gas acceleration directly traces the net forces acting on the Interstellar Medium (ISM). It is therefore a powerful diagnostic for identifying and quantifying the drivers of gas dynamics, revealing how gas responds to gravity, radiation pressure, magnetic fields, and stellar feedback. However, measuring acceleration in the ISM is challenging. Spectroscopic line observations provide kinematic information in Position-Position-Velocity (PPV) space, leading to ambiguities, known as PPV degeneracy: different parts of a cloud, located at different physical distances or moving in different directions, can appear at the same position in the sky and share similar velocities, making their true spatial configuration and true motion ambiguous \citep[e.g.,][]{beaumont_quantifying_2013}. Breaking this degeneracy and inferring acceleration requires knowledge of the three-dimensional physical space, which is necessary to deproject the PPV data.

Acceleration in the ISM has so far been recovered only indirectly, by modeling the projection of line profiles, gas kinematics \cite[e.g.,][]{hacar_gravitational_2017}, or stellar population motions. In Sco-Cen \cite[e.g.,][]{ratzenbock_star_2023,ratzenbock_significance_2023,hutschenreuter_velocity_2026}, for example, young clusters trace $\sim$100~pc-long sequences accelerating outward from the central OB groups at $\sim$0.4--0.8~km/s/Myr \citep{posch_corona_2023, posch_physical_2025, miret-roig_tw_2025,grosschedl_evolution_2026}, a clean fingerprint of past feedback recorded in the stellar component. The gas component has never been measured this way: the line-of-sight projection of the velocity field cannot be inverted without independent three-dimensional information. Recently, 3D dust maps based on Information Field Theory \citep{enslin_information_2009} applied to Gaia data \citep{leike_resolving_2020,edenhofer_parsec-scale_2024} resolve nearby molecular cloud complexes at parsec scales and yield robust inclinations for elongated structures \citep{zucker_three-dimensional_2021, Zucker_PPVII}, supplying the geometry needed to invert a line-of-sight velocity field \citep{soding_bayesian_2023,konietzka_radcliffe_2024,soler_kinetic_2025}.

Filament formation scenarios offer a natural and powerful testing ground for gas dynamics (see reviews in \citealt{hacar_initial_2023}; \citealt{pineda_bubbles_2023}). Specifically, the recently proposed model of R-type filaments posits that stellar feedback from OB stars compresses remaining dense gas and naturally stretches it away from the source of feedback, resulting in a distinct head-to-tail morphology \citep{alves_hp2_2025}. This scenario predicts a unidirectional, longitudinal flow away from the feedback source. The B44 filament in Ophiuchus (see Figure~\ref{fig:herschel}) is a prime example of such an R-type structure, with its star-forming head facing the Upper-Sco OB association. Testing this prediction requires measuring the true 3D flow of gas inside this filament.

In this paper, we find that gas flows longitudinally along B44, accelerating away from the Upper-Sco OB association. This result matches the kinematic signature predicted for an R-type filament, and it shows that Gaia 3D dust maps can lift the line-of-sight ambiguity intrinsic to PPV spectral data, allowing the gas velocity field in coherent filaments to be deprojected directly. Furthermore, our analysis places B44 within a cooling-driven entrainment regime, linking the current gas dynamics to the long-term feedback history of the entire Upper-Sco complex.

\begin{center}
\begin{figure*}[!tbp]
\includegraphics[width=\hsize]{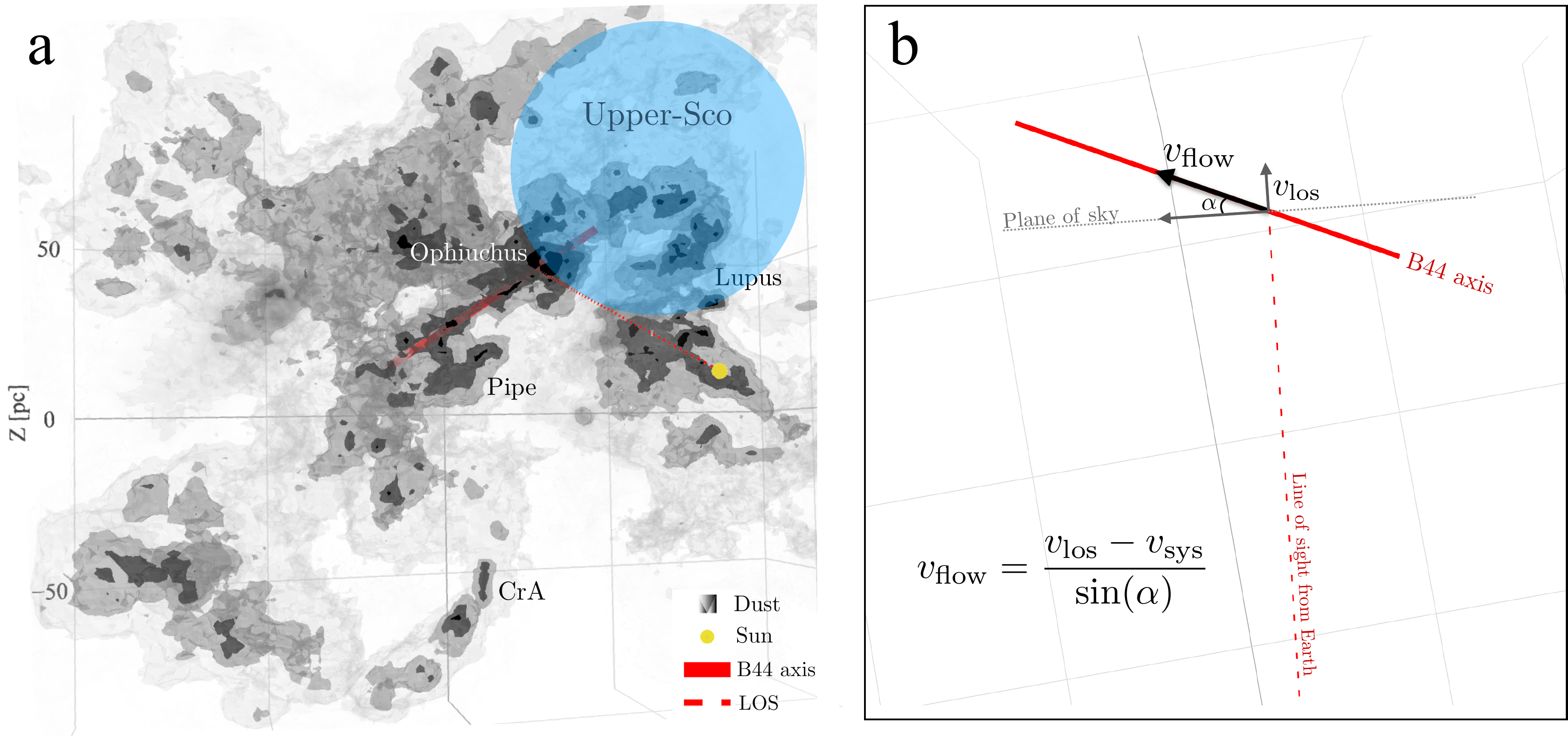}
\caption{(a) 3D spatial context of the B44 filament, relative to nearby molecular clouds (using the \citealt{leike_resolving_2020} 3D dust map). The solid red line traces the axis of the B44 filament while the dashed red line indicates the line of sight from Earth. The general location of the ionizing OB stars in Upper-Sco is marked as a transparent light-blue circle. For a better visualization, see the interactive version of this figure \href{https://faun.rc.fas.harvard.edu/czucker/Paper\_Figures/B44\_Flow.html}{here}. A permanent copy is archived at \href{https://doi.org/10.5281/zenodo.21498960}{Zenodo}. (b) Top-down view: schematic illustration of the geometry used to relate the observed line-of-sight velocity component $v_{\mathrm{los}}$, or radial velocity, to the true velocity of gas flow $v_{\mathrm{flow}}$. Knowledge of the angle $\alpha$ between the filament axis and the plane of the sky allows one to infer $v_{\mathrm{flow}}$ directly.}
\label{fig:3D}
\end{figure*}
\end{center}

\section{Data and results}

To measure the three-dimensional gas flow along B44, we combine two datasets: the 3D dust map of \cite{leike_resolving_2020}, which constrains the spatial orientation of the filament, and the COMPLETE Survey $^{12}$CO(1-0) and $^{13}$CO(1-0) maps of \citet{ridge_complete_2006}, which provide the line-of-sight velocity field. To follow the flow beyond the region mapped by COMPLETE, we additionally use the lower-resolution $^{12}$CO survey of \citet{dame_milky_2001}, which images the entire Ophiuchus complex. This dataset lets us trace the same filament axis well past L1729 and test whether the velocity gradient continues to steepen or instead saturates (Sect.~\ref{sec:dame}, Appendix~\ref{app:dame}).

\subsection{Filament geometry and deprojection}

We determined the spatial orientation of B44 by fitting a line in three dimensions to the spine of the filament in the \cite{leike_resolving_2020} dust cube. We apply a Principal Component Analysis (PCA) to the 3D voxels tracing the B44 filament and weight the PCA by the hydrogen nuclei volume density of each voxel, restricting the fit to denser voxels $n > 30 \; \mathrm{cm}^{-3}$, adopting the dust-to-gas ratio described in \cite{zucker_three-dimensional_2021}.  We then compute the angle between the first principal component and the plane of the sky. A full description of the methods underpinning the PCA analysis is provided in Appendix~\ref{app:pca}.

The resulting inclination angle is $\alpha = 13^\circ \pm 2^\circ$, where the central value is derived from applying the PCA to the mean 3D dust map of \cite{leike_resolving_2020}, and the uncertainty is given by the standard deviation of $\alpha$ obtained by repeating the same analysis on each of the 12 posterior realizations of the dust map. Figure~\ref{fig:3D} illustrates this geometry, showing the 3D dust map of Ophiuchus with the B44 axis (solid red line) and the LOS to L1689 (dashed red line). The geometry is more clearly seen in the accompanying 3D interactive visualization. The assumption of a single inclination angle is supported by the morphology of B44, which appears as a coherent, approximately linear structure both in Herschel column density maps (Figure~\ref{fig:herschel}) and in the Gaia-based 3D dust reconstruction. We verified this directly in three dimensions, which is most readily done in the interactive version of Figure~\ref{fig:3D}: B44 is straight over its full length, with no appreciable change in the direction of its axis between L1689 and L1729 that would make $\alpha$ vary with position along the filament. Although the more recent map of \cite{edenhofer_parsec-scale_2024} is also available, its spherical reconstruction grid could affect the inferred orientation of the filament, so we adopt the Cartesian-grid map of \cite{leike_resolving_2020}.

With $\alpha$ in hand, we can deproject both the line-of-sight velocities and the projected separations on the sky. The observed line-of-sight velocity traces only the component of gas motion projected along the LOS, so the true flow velocity along the filament axis is
\begin{equation}
    v_{\rm flow} = \frac{v_{\rm los} - v_{\rm sys}}{\sin (\alpha)}
    \label{eq:vflow}
\end{equation}
where $v_\mathrm{sys}$ is the systemic velocity of the cloud. Similarly, the projected separation on the sky ($d_\mathrm{sky} \approx 8$~pc) is deprojected to the true filament length $\Delta s = d_\mathrm{sky}/\cos\alpha \approx 8$~pc (with $\alpha = 13^\circ$ the deprojection factor $1/\cos\alpha \approx 1.03$ is nearly unity). Given the small angular extent of B44 from L1689 to L1729 ($\Delta\theta < 3^\circ$), the projection angle $\alpha$ remains effectively constant along the filament, so deprojection reduces to a single global factor and the observed velocity gradient directly traces the physical acceleration without position-dependent corrections. This deprojection assumes that the gas motion is predominantly longitudinal, an assumption we test against rotation and other kinematic origins in Sect.~\ref{sec:alternatives} before interpreting the gradient as acceleration.

\begin{figure*}[!tbp]
\includegraphics[width=\hsize]{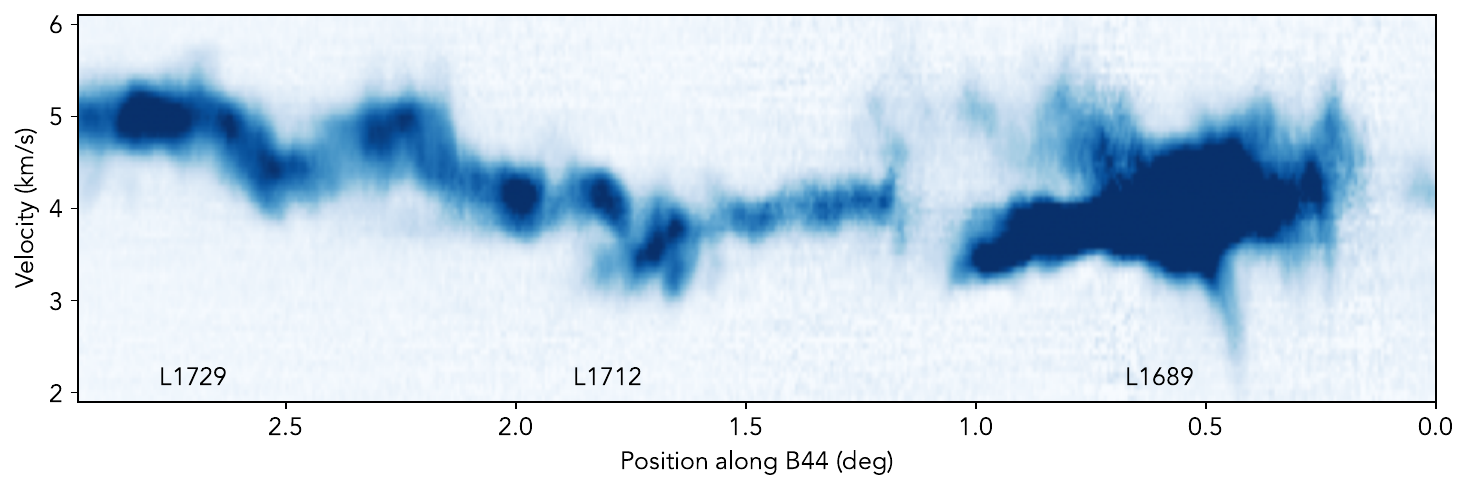}
\caption{Position-velocity (LSR) diagram of $^{13}$CO along the B44 filament axis, from L1689 (head of B44) to L1729 (tail). The overall velocity increases from L1689 to L1729, tracing the large-scale longitudinal flow discussed in this work. Superimposed on this trend are abrupt $V_\mathrm{LSR}$ steps, or jumps, between individual clumps, as first reported by \citet{loren_cobwebs_1989}, which we interpret as the natural outcome of the interaction between the low-density feedback flow from Upper-Sco and dense cores of varying mass and cross-section. Narrow, nearly vertical features visible at several positions along the filament likely trace localized shocks produced by this same interaction, where the impinging flow compresses the molecular gas and produces emission over a broad velocity range at a single spatial location. That this PV substructure is distributed along the filament rather than confined to the head is consistent with an interaction between the flow and the dense gas that occurs along the full length of the filament (Appendix~\ref{sec:shocks}).}
\label{fig:PV13}
\end{figure*}

\subsection{Velocity field along B44}
\label{sec:velfield}

\begin{figure}
    \centering
    \includegraphics[width=\linewidth]{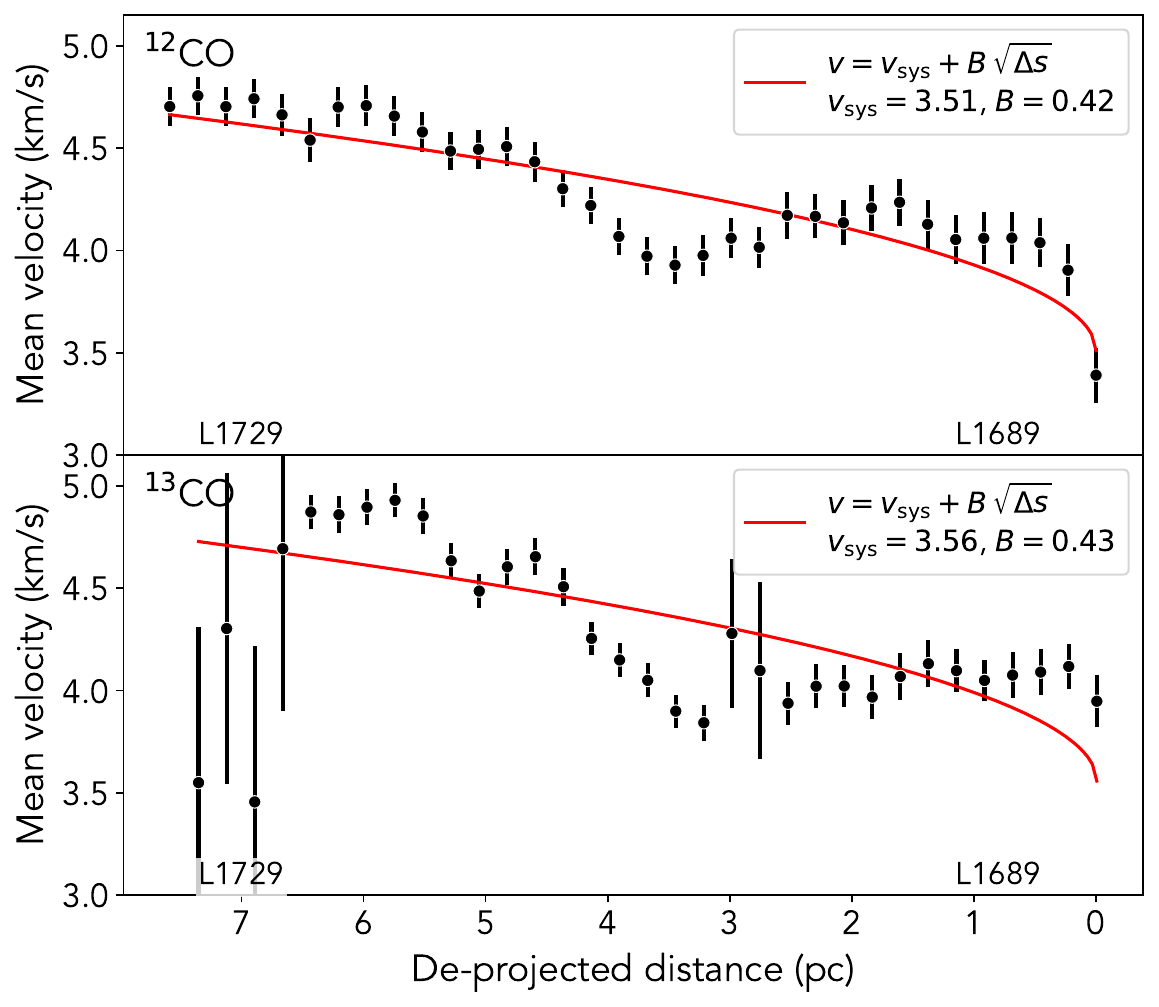}
\caption{Mean velocity along the B44 filament as a function of deprojected distance from L1689, shown separately for $^{12}$CO (top) and $^{13}$CO (bottom), with error bars showing the standard error on the mean (SEM) in each spatial bin. The red curves show a constant-acceleration model $v = v_\mathrm{sys} + B\sqrt{\Delta s}$, fitted to the data weighted by the SEM.
The $^{12}$CO fit yields $v_\mathrm{sys} = 3.51$~km/s and $B \approx 0.42$~km/s/pc$^{1/2}$, corresponding to a mean acceleration of $a = B^2/(2\sin^2\alpha) \approx 1.8$~km/s/Myr. The independent $^{13}$CO fit gives consistent values ($v_\mathrm{sys} = 3.56$~km/s, $B \approx 0.43$~km/s/pc$^{1/2}$). The model has an intrinsically steep slope near $\Delta s = 0$ that is not well constrained by the data, so the fitted $v_\mathrm{sys}$ is sensitive to the behavior near the head, and the derived acceleration is best read as a characteristic value averaged over the observed length of the filament (Appendix~\ref{app:caveats}).}
\label{fig:PV12_fit}
\end{figure}

The COMPLETE Survey $^{12}$CO(1-0) and $^{13}$CO(1-0) maps cover the head of the filament (L1689) down to L1729 (see Figure~\ref{fig:herschel}), with sufficient spectral resolution to resolve the velocity structure along the filament. To measure the velocity field along the filament axis, we computed intensity-weighted mean velocities in 40 spatial bins tilted to align with the filament orientation on the sky and spanning its full observed extent (bin width $\approx 0.22$~pc on the plane of the sky, $\approx 0.23$~pc once deprojected with $\alpha = 13^\circ$). In each bin, the mean velocity is $v_{\rm mean} = \sum I(v)\,v / \sum I(v)$, where $I(v)$ is the summed intensity across all spatial pixels in the bin at velocity channel $v$. We iteratively rejected velocity channels lying more than $5\sigma_v$ from the current mean, where $\sigma_v$ is the intensity-weighted velocity dispersion of the bin (not the antenna-temperature noise), recomputed at each step and with a maximum of 10 iterations, before computing the final mean and dispersion. The reported RMS ($\sim$0.5--0.9~km/s) is the intensity-weighted velocity dispersion of the summed spectrum in each bin; in an optically thick tracer such as $^{12}$CO this quantity mixes the bulk-motion dispersion of centroids across pixels with opacity-broadened linewidths and is therefore not a clean measure of either \citep[see, e.g.,][]{hacar_opacity_2016}. The statistical uncertainty on the mean is the standard error on the mean, SEM$~\approx 0.1$~km/s, which we adopt as the error bar in all subsequent fits. The $^{12}$CO and $^{13}$CO velocity profiles along B44 agree closely (Sect.~\ref{sec:accel}, Figure~\ref{fig:PV12_fit}). The two lines differ in optical depth by more than an order of magnitude, so an artifact of opacity or excitation would be unlikely to reproduce itself in both, and their agreement indicates that the gradient is kinematic in origin.

Figure~\ref{fig:PV13} shows the resulting position-velocity diagram, and Figure~\ref{fig:PV12_fit} shows the intensity-weighted mean velocities as a function of deprojected distance from L1689. The mean $^{12}$CO velocity increases from $\sim$3.5~km/s at L1689 (the star-forming head) to $\sim$4.7~km/s at L1729 (the tail), a gradient of $\sim$1.2~km/s over $\sim$8~pc in deprojected distance.

\subsection{A longitudinal flow, not rotation}
\label{sec:alternatives}

For the monotonic $v_\mathrm{los}$ gradient to trace genuine longitudinal acceleration, rather than rotation, transverse motion, or gravitational streaming, the bulk gas motion must be directed along the filament axis and driven by the Upper-Sco wind. Over the $\sim$8~pc covered by the COMPLETE maps this cannot be settled from the kinematics alone: an accelerating-flow law $v = v_\mathrm{sys} + B\sqrt{\Delta s}$ and a solid-body-rotation law $v = v_0 + k\,\Delta s$ fit the gradient about equally well over that baseline. We break this degeneracy with the velocity profile traced beyond L1729, and then address the remaining alternatives in turn.

\begin{figure}
    \centering
    \includegraphics[width=\linewidth]{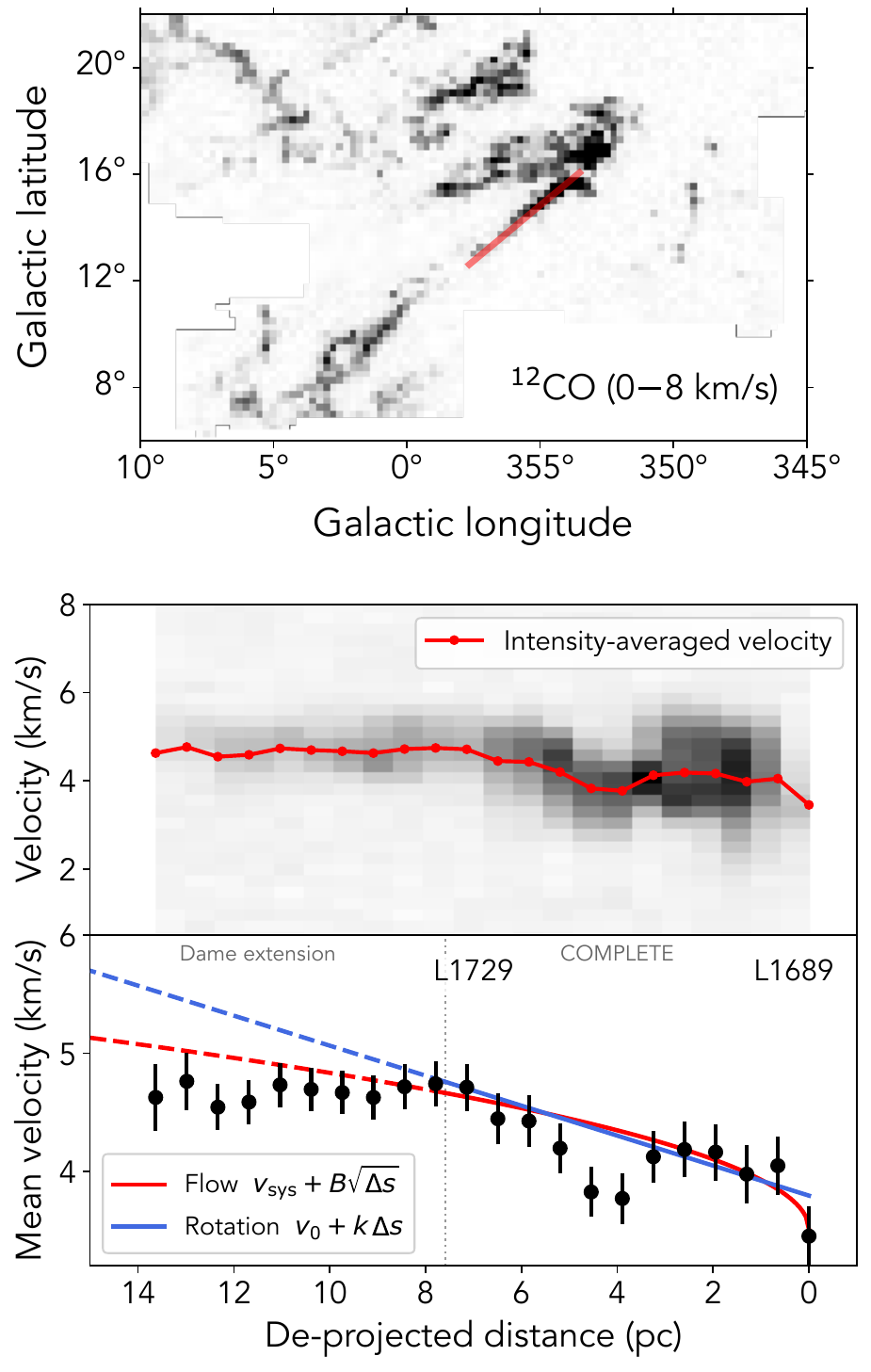}
    \caption{Extended kinematics of B44 from the \citet{dame_milky_2001} $^{12}$CO survey. Top: integrated CO emission (0--8~km/s) over the Ophiuchus complex with the position--velocity cut drawn as the red swath (the head L1689 lies at upper right). Middle: position--velocity slice along the cut (head on the right) with the intensity-weighted mean velocity in red. Bottom: mean velocity versus deprojected distance from L1689 (black points), with the COMPLETE constant-acceleration flow (red) and solid-body rotation (blue) fits of Fig.~\ref{fig:PV12_fit} extrapolated beyond their fitted range (dashed). The gradient continues past L1729 and flattens into a plateau, falling below both extrapolations and furthest below the linear rotation law.}
    \label{fig:dame}
\end{figure}

\emph{The extended velocity profile.}\label{sec:dame} The COMPLETE maps cover B44 only to $\sim$8~pc from the head, but the lower-resolution CO survey of \citet{dame_milky_2001}, which images the whole Ophiuchus complex, lets us follow the same axis well past L1729. Using the identical spine and procedure as in Sect.~\ref{sec:velfield} (Appendix~\ref{app:dame}), we recover the profile in Fig.~\ref{fig:dame}. The head-to-tail gradient is reproduced in this independent, lower-resolution dataset and continues beyond L1729, but instead of steepening it flattens into a plateau at $v_\mathrm{los} \approx 4.7$~km/s by a deprojected distance of $\sim$9~pc. This is the decisive kinematic discriminator: an accelerating flow approaching a terminal velocity levels off in just this way, whereas solid-body rotation would keep steepening linearly with no turnover. The two laws diverge precisely here, and the data follow the saturating flow, falling below both extrapolations and furthest below the linear rotation law (Fig.~\ref{fig:dame}). The physical origin of the plateau, which sits well below the feedback-wind speed, is discussed in Sect.~\ref{sec:feedback_flow}. We caution that the Dame cube resolves only $\sim$0.6~pc per pixel and traces $^{12}$CO alone, so although the plateau itself is robust (Appendix~\ref{app:dame}), its precise saturation scale is uncertain.

\emph{Transverse motions.} Accretion onto the filament, sheared flows, or a wind misaligned with the filament axis would all add a transverse component to $v_\mathrm{los}$ without producing a true longitudinal acceleration. The morphology, on the other hand, argues against transverse: B44 is a coherent, approximately linear filament of essentially constant width (1--2~pc) over its full length, in both the Herschel column density map (Fig.~\ref{fig:herschel}) and the \citet{leike_resolving_2020} 3D dust reconstruction, with no curvature or width gradient, whereas strong transverse motions or cross-winds would shear or broaden it. The close agreement of the $^{12}$CO and $^{13}$CO profiles, which sample different optical depths, further argues against a tracer- or geometry-specific artifact. Upper-Sco does contain $\sim$20 distributed massive stars \citep{alves_hp2_2025}, so some cross-wind component is expected, but the constant width indicates a single dominant wind direction approximately aligned with the filament, with cross-flows only a subdominant perturbation. Any residual transverse component is carried as a systematic in Appendix~\ref{app:caveats}. We note that the plane-of-sky motion of the gas cannot be measured directly here. Gaia delivers precise astrometry for point sources, but a diffuse, extended structure such as B44 lacks the compactness that an astrometric solution requires, so cloud motions have conventionally been inferred from proxies, most often the proper motions of embedded young stars \citep[e.g.,][]{grosschedl_3d_2021}. That proxy is weak toward B44: the filament is opaque enough that few background stars are seen through it, and its young stellar population is concentrated at the star-forming head, while the body and tail are essentially starless, meaning, no stellar tracer samples the length of the flow. Direct measurement of interstellar cloud proper motions has recently been demonstrated using image registration on multi-epoch near-infrared imaging \citep{piecka_direct_2025}, which offers a promising route to the transverse component of the B44 flow in future work.

\emph{Gravitational streaming.} \citet{loren_cobwebs_1989} searched this system for streaming motions and found only abrupt clump-to-clump $V_\mathrm{LSR}$ jumps. We confirm these jumps (Fig.~\ref{fig:PV13}) but read them as differential entrainment of clumps by the feedback flow ($v \sim 20$~km/s; \citealt{piecka_towards_2024}) rather than self-gravitating streaming: collapse or wave modes \citep[e.g.,][]{gehman_linear_1996,hacar_dense_2011} would drive gas toward the densest part of the filament or produce sign changes in the profile, not the monotonic head-to-tail increase observed here.

\subsection{Gas acceleration}
\label{sec:accel}

Having established that the gradient traces a longitudinal flow rather than rotation or transverse motion (Sect.~\ref{sec:alternatives}), we now measure its acceleration. We model B44 as gas set in motion from rest at the head and accelerated uniformly along the filament by a steady force. Under this assumption the flow speed grows with distance as $v_\mathrm{flow}^2 = 2a\,\Delta s$, where $\Delta s$ is the distance from L1689 along the filament axis and $a$ is the mean acceleration. Projecting this flow onto the line of sight through the inclination angle $\alpha$ gives the observed velocity $v_\mathrm{obs} = v_\mathrm{sys} + \sin\alpha\,\sqrt{2a\,\Delta s} \equiv v_\mathrm{sys} + B\sqrt{\Delta s}$, with $B = \sin\alpha\,\sqrt{2a}$ and $v_\mathrm{sys}$ the systemic velocity of the cloud. We fit this square-root law to the full deprojected velocity profile (Figure~\ref{fig:PV12_fit}) rather than differencing the endpoints or relying on a single spatial bin, because it determines $v_\mathrm{sys}$ and the acceleration simultaneously from the shape of the entire profile and is therefore less sensitive to local perturbations near the wind-facing head (Appendix~\ref{app:caveats}).
The fit, weighted by the standard error on the mean in each bin, yields $v_\mathrm{sys} = 3.51$~km/s and $B \approx 0.42$~km/s/pc$^{1/2}$ for $^{12}$CO, with an independent $^{13}$CO fit giving consistent values ($v_\mathrm{sys} = 3.56$~km/s, $B \approx 0.43$). Inverting $B = \sin\alpha\,\sqrt{2a}$ gives the mean acceleration \begin{equation} a = \frac{B^2}{2\sin^2\alpha} \approx 1.8~\mathrm{km/s/Myr} \approx 6 \times 10^{-11}~\mathrm{m/s}^2 \label{eq:accel} \end{equation} As a check that the inferred acceleration does not depend on the specific functional form, we also fitted a linear model $v_\mathrm{obs} = v_0 + k\,\Delta s$ to the same data and evaluated the equivalent mean acceleration $\langle a \rangle = (k/\sin\alpha)^2\,L/2$, where $L$ is the profile length. This yields $\langle a \rangle \approx 1.2$~km/s/Myr for $^{12}$CO and $\langle a \rangle \approx 1.7$~km/s/Myr for $^{13}$CO, in agreement with the square-root fit within the Monte Carlo uncertainty.

To estimate the uncertainty, we performed a Monte Carlo analysis with $10^5$ realizations (see Appendix~\ref{app:errors} and Figure~\ref{fig:mc_errors}). The resulting acceleration for $^{12}$CO is $a = 1.77^{+0.74}_{-0.47}$~km/s/Myr (16th--84th percentile range). The $^{13}$CO analysis gives $a = 1.88^{+0.79}_{-0.51}$~km/s/Myr, which agrees with the $^{12}$CO value well within the uncertainties.

\section{Discussion}                                                                  
                                                                       
\subsection{Evidence for feedback-driven longitudinal flow}
\label{sec:feedback_flow}
The velocity gradient measured along B44 runs from the head (L1689), which faces the massive stars in Upper-Sco, toward the tail (L1729), with gas accelerating away from the OB association. This is the kinematic signature predicted by the R-type filament scenario of \citet{alves_hp2_2025}, where stellar feedback stretches dense gas pockets into elongated head--tail structures. The direction of the flow is notable: gas moves away from the dense, star-forming head toward the diffuse tail. This is opposite to the classical hub-filament picture, in which gas flows along filaments toward a central hub \citep[e.g.,][]{myers_filamentary_2009,peretto_sdc13_2014}. In B44, the dense, star-forming head is the origin of the flow, not its destination. To our knowledge, this is the first inference of three-dimensional gas acceleration inside a molecular cloud filament based on a measured filament inclination. 

A simple momentum-flux estimate using fiducial values for the Upper-Sco feedback flow ($\rho \sim 10$~cm$^{-3}$, $v \sim 20$~km/s; \citealt{piecka_towards_2024}), coupled to a filament cross-section of $\sim 1$~pc$^2$ and mass of the dense gas at the head of the filament $\sim 100$~M$_\odot$, yields a characteristic acceleration of $\sim 1$~km/s/Myr, in rough agreement with the observed value given the uncertainties on the density of the impinging medium (see Appendix~\ref{app:momentum} for details). This figure characterizes the momentum budget available to drive the flow, but it does not by itself specify the coupling mechanism.

The same massive stars also radiate, so radiative feedback deserves consideration as an alternative driver, through either radiation pressure or photoevaporation. Radiation pressure is straightforward to bound. Upper-Sco contains 20 stars of spectral type B3 or earlier, with a combined luminosity of order $10^5$~L$_\odot$ dominated by the few earliest types \citep{alves_hp2_2025}. At the $\sim$10~pc separation of B44, and assuming the filament absorbs all the radiation incident on it, the intercepted momentum flux $L A / (4 \pi d^2 c)$ spread over the same $\sim$100~M$_\odot$ of dense head gas gives $\sim$0.02~km/s/Myr, two orders of magnitude below the observed acceleration and far weaker than the wind term. The estimate scales linearly with $L$ and as $d^{-2}$, so the conclusion is not sensitive to the exact luminosity adopted. Photoevaporation is excluded observationally rather than energetically. A rocket-effect recoil requires an ionization front at the wind-facing surface, which would appear as an H$\alpha$ rim and free-free emission and would leave B44 with the bright-rimmed, cometary morphology characteristic of radiation-driven implosion. No such rim is seen at the head of B44 or along its length, and the head is a dark, dense clump. The nearest ionizing star, HD~147889 ($Q_\mathrm{H} \approx 10^{47.4}$~s$^{-1}$; \citealt{alves_hp2_2025}), powers a photodissociation region on the western face of L1688, that is, a neutral FUV-heated layer rather than an ionization front driving into the filament. We therefore find no indication that radiation contributes appreciably to the acceleration measured here. 

The shape of the velocity profile, a quasi-flat plateau near the head followed by a smooth rise toward the tail (see Appendix~\ref{app:caveats}, Table~\ref{tab:subrange}), is consistent with the two-phase entrainment picture of \citet{gronke_how_2020}, in which a cold cloud first sheds material from its wind-facing surface during a ``tail-formation'' phase and then progressively grows in mass as the stripped gas mixes, recools, and accumulates downstream. Assuming this picture scales down to the B44 case, gas at larger downstream distances has spent progressively longer in the flow and will tend to move faster than gas upstream.

The relevant picture is the cooling-driven entrainment of cold gas in a hot wind \citep{klein_hydrodynamic_1994, scannapieco_launching_2015, gronke_growth_2018, gronke_how_2020}: material stripped from the wind-facing surface mixes with the impinging flow, and, provided the mixed-gas cooling time is shorter than the cloud-crushing time ($t_\mathrm{cool,mix}/t_\mathrm{cc} \lesssim 1$), recools onto the cold phase and is advected downstream rather than mixing away. With fiducial parameters for B44 this ratio of timescales is formally $\sim$$10^{-3}$--$10^{-2}$, although B44 lies in a much colder and lower-velocity regime than the simulations that calibrated this criterion, so we adopt the entrainment picture only qualitatively. The mechanism and the applicability of this framework to B44 are discussed in Appendix~\ref{app:entrainment}. In this picture the L1689 head plays the role of the cold cloud, while L1712 and L1729 trace the entrained, downstream tail formed from stripped and recooled material; the present-day cylindrical morphology of B44 is the shaped outcome of this process, not its initial condition. The coupling therefore proceeds through entrainment and ablation at the wind-facing head (L1689), not through bulk compression of the filament.

The lower-resolution Dame CO survey offers a first, if tentative, test of this picture (Sect.~\ref{sec:dame}). Beyond the COMPLETE coverage, past L1729, the velocity does not keep rising but levels off into a plateau at only about a quarter of the $\sim$20~km/s feedback wind. Rather than the gas approaching the wind speed, this is the behavior expected if the flow coasts as the Upper-Sco ram pressure dilutes with distance from the source over a roughly constant gas column \citep{alves_hp2_2025}. We treat this as suggestive rather than established, since it rests on $^{12}$CO data at $\sim$0.6~pc resolution and will require higher-resolution radio mapping to confirm.

\subsection{Comparison with other feedback environments}

Beyond B44, velocity gradients associated with stellar feedback have been observed in other star-forming environments, providing useful context for our measured acceleration.
\citet{tachihara_molecular_2000} identified streaming motions in gas near the runaway O star $\zeta$~Oph, interpreted as evidence of the dynamical influence of massive stars on the surrounding ISM. High-resolution mapping of the L260 end of L204, one of the T-type filaments identified in \citet{alves_hp2_2025}, reveals a velocity gradient oriented perpendicular to the filament's long axis \citep{gong_widespread_2022}, the transverse counterpart to the longitudinal flow we measure in the radial case.
\citet{dent_large-scale_2009} find evidence of gas acceleration primarily through velocity gradients in compact molecular clumps near the Rosette Nebula's O stars, assuming an inclination of the clumps. Nevertheless, they derived accelerations about an order of magnitude higher than in this work, although it is likely due to the substantially different star-forming environment, containing an order of magnitude larger number of O-stars than Upper-Sco.

\subsection{A bow shock at the head of B44}
\label{sec:bowshock}

If the supersonic Upper-Sco feedback flow impinges on the dense head of B44, it should drive a bow shock at L1689, and a bow-shaped morphology is indeed seen in the \textit{IRAS} 100\,$\mu$m emission, opening downstream as expected when a fast flow is deflected around a dense obstacle \citep[see Figure~8 of][]{alves_hp2_2025}. The opening angle implies a flow Mach number $\mathcal{M} \approx 2$, but referred to the sound speed of the upstream wind, whose temperature is poorly constrained. We therefore treat the morphology as qualitative evidence for a shock and anchor its strength in the post-shock density jump, a quantity internal to the cold molecular gas.

A cut along the filament axis in the Herschel column-density map, running from the undisturbed gas upstream of L1689 through the head and down the body, shows a peak-to-upstream contrast of $\approx 5$. For an isothermal shock the compression sets the Mach number through $\rho_{\rm post}/\rho_{\rm pre} \approx \mathcal{M}^2$, so this contrast inverts to
\begin{equation}
\mathcal{M}_{\rm shock} = \sqrt{\rho_{\rm post}/\rho_{\rm pre}} \approx \sqrt{5} \approx 2.2,
\end{equation}
now referred to the measured cold-gas sound speed $c_{\rm s} \approx 0.21$\,km\,s$^{-1}$ (from the NH$_3$ kinetic temperature $T_{\rm K} = 12.5$\,K at L1689 and L1712; \citealt{friesen_green_2017}). This is a lower bound: the contrast is measured in column density, not volume density, which should push the true $\mathcal{M}_{\rm shock}$ down. The bow morphology and the density jump thus agree on a mildly supersonic shock, $\mathcal{M} \approx 2$, with the density-jump value the more reliable because its reference sound speed is measured rather than assumed. The same cut shows the column density falling off toward the body and tail, so the mass is concentrated at the wind-facing head, as assumed in our momentum budget (Appendix~\ref{app:momentum}).

The post-shock gas shows the sonic transition expected behind a shock: the $^{13}$CO envelope retains transonic turbulence ($\sigma_{\rm NT} \approx 0.34$\,km\,s$^{-1}$, $\mathcal{M}_{\rm turb} \approx 1.6$), while the dense NH$_3$ spine has relaxed to subsonic, coherent motion ($\sigma_{\rm NT} = 0.17$\,km\,s$^{-1}$, $\mathcal{M}_{\rm turb} \approx 0.8$; \citealt{pineda_direct_2010, hacar_cores_2013}). The head itself splits into two sub-condensations (L1689N and L1689S) at the bow apex, as seen in simulations of shocked clumps \citep{goldsmith_interaction_2016, goldsmith_comparison_2018}, and further shock signatures appear in the position-velocity structure along the filament (Appendix~\ref{sec:shocks}).

\subsection{Timescale and steady-state flow}

The measured flow velocity and acceleration within B44 set a kinematic timescale for the flow. Assuming a constant acceleration of $a \approx 1.8$~km/s/Myr, a gas parcel launched from rest at the head reaches a distance $\Delta s$ after a time $t = \sqrt{2\Delta s/a}$. For the observed extent of $\Delta s \approx 8$~pc, this gives $t \approx 3.0$~Myr, consistent with the independent estimate $t = v_\mathrm{flow}/a \approx 5.3/1.8 \approx 3.0$~Myr. This figure is well within the $\sim$10--11~Myr age of the Upper-Sco \citep[e.g.,][]{ratzenbock_star_2023}, indicating that the feedback flow has been present long enough to establish the flow we observe today.

We emphasize that this $\sim$3~Myr figure is the kinematic age of the flow, not the lifetime of a coherent filamentary structure. This kinematic age likely exceeds the transit time of any individual gas parcel through the observed extent.  The morphology we see is maintained by the balance between ablation at the wind-facing head and downstream advection, not by the longevity of any particular gas parcel. In this picture, the lifetimes of individual coherent substructures within B44 (clumps, cores) can remain short, in the $\sim$1~Myr range typical of molecular gas \citep[e.g.,][]{hacar_initial_2023}, while the large-scale flow pattern persists on the $\sim$10~Myr feedback timescale. 
The system has therefore been exposed to the flow long enough that a dynamically maintained, quasi-steady head-to-tail morphology is plausible. 

Extrapolating the constant-acceleration model, the kinematic time to reach 15~pc is $\sim$4.1~Myr. These timescales remain comfortably within the age of Upper-Sco. This suggests that B44 is still actively growing, with the feedback flow continuing to stretch the filament and feed gas from the head toward an ever-receding tail. The lower-resolution Dame CO survey already provides a first test of this prediction (Sect.~\ref{sec:dame}): the gradient levels off beyond $\sim$9~pc rather than continuing to rise, so the extrapolated velocities above are upper limits. Deeper, higher-resolution observations would pin down the saturation scale.

\subsection{Connection to the Sco-Cen cluster-chains}

The B44 flow provides a present-day, gas-phase counterpart to the fossil record preserved in the Sco-Cen cluster chains. \citet{posch_corona_2023, posch_physical_2025} identified three $\sim$100~pc-long sequences of clusters in Sco-Cen, and measured accelerations of $\sim$0.4--0.8~km/s/Myr along them. 

B44 captures the same process in its currently active phase: gas that has not yet collapsed into stars, still being pushed and stretched by the same Upper-Sco feedback flow that shaped the chain of clusters ending on it. The acceleration we measure in B44's gas today ($\sim$1.8~km/s/Myr) is some three times larger than the rates along the Sco-Cen cluster-chains. Three effects plausibly account for a difference of this size. The chains record a 15-Myr time-average, whereas B44 records the current rate. The wind may have strengthened recently. And B44 simply lies closer to the source of feedback than the Corona and LCC chains do. On any of these readings the process is the same one, and B44 is its gas-phase extension, caught mid-formation. The gas mass of B44 is $\sim$1500~M$_\odot$, with $\sim$1000~M$_\odot$ above A$_\mathrm{K}$ of 0.8~mag (see Table 1 in \citealt{alves_hp2_2025}), so it is a substantial reservoir of star-forming gas that could form a new generation of stars in the same $\rho$-Oph chain \citep{posch_physical_2025}. Assuming an efficiency of $\sim$30\% \citep[e.g.,][]{alves_mass_2007,lada_star_2010}, the B44 cloud will produce 600--700 stars.

\subsection{A possible role for the magnetic field}
\label{sec:bfield}

A magnetic field is likely present in B44. The optical polarization mapping of \cite{goodman_optical_1990} shows that the elongated clouds, in particular B45, tend to align at an intermediate angle set by their long axis and the local large-scale field, rather than strictly parallel or perpendicular to it. \cite{goodman_magnetic_1994} used the Zeeman effect in 21 cm H I self-absorption to measure a mean uniform line-of-sight field of about 10.2 $\mu$G near L1688.  At the head of B44, \citet{pattle_jcmt_2021}, mapping 850~$\mu$m dust polarization in L1689 with JCMT POL-2 as part of the BISTRO survey, inferred plane-of-sky field strengths of $\sim$70--370~$\mu$G across its dense clumps (L1689N, SMM-16, and L1689B) using the Davis-Chandrasekhar-Fermi (DCF) framework \citep{1951PhRv...81..890D,1953ApJ...118..113C}. They found these regions to be magnetically transcritical with sub-Alfv\'enic turbulence. In L1689 the field is ordered, runs approximately perpendicular to the local filament, and is consistent with the cloud-scale field traced by Planck \citep{2015A&A...576A.104P}.  On core scales the picture is more complex, since ongoing gravitational collapse may confuse the geometry, but the magnetic field direction is still largely consistent with larger scales.

The field may also shape the flow, in two ways that we note without endorsing either. A field threading the tail could collimate and contain the entrained gas, its tension resisting lateral dispersal and channeling stripped material along the field lines. This would offer a natural contribution to the nearly constant width of B44 over its full length (Sect.~\ref{sec:alternatives}). 

A more specific and frankly speculative possibility is that B44 streams away from Upper-Sco while rotating about its own long axis. \citet{uchida_velocity_1990} proposed exactly such a configuration for the neighboring B45 streamer, interpreting a velocity split along it as a spinning streamer that drains angular momentum from L1688 and thereby promotes star formation there. The Herschel column density map of B44 (see Figure~\ref{fig:herschel}) shows hints of a helical or double-spiral morphology between L1689 and L1712, and preliminary polarization measurements of the field geometry point in the same direction (Y.~Matsumura et al. 2016, unpublished). We stress that a torsional spin about the long axis is distinct from the transverse solid-body rotation excluded in Sect.~\ref{sec:alternatives}: it would not produce a head-to-tail $v_\mathrm{los}$ gradient and is fully compatible with the longitudinal flow measured here. Testing this picture will require sensitive dust-polarization mapping and a search for a systematic velocity twist across the filament, which we leave to future work.

\section{Conclusions}

Gas in B44 flows longitudinally from its star-forming head (L1689) toward its tail (L1729), accelerating away from the massive stars of Upper-Sco at approximately \SI[per-mode=symbol]{1.8}{\kilo\metre\per\second\per\mega\year} ($\sim \SI{6e-11}{\metre\per\second\squared}$). The measurement rests on deprojecting the observed radial velocities through an inclination fixed independently by the Gaia-based 3D dust map. The two CO lines, which differ substantially in optical depth, give consistent profiles and accelerations, so the gradient is a genuine kinematic signal (Sect.~\ref{sec:velfield}). In the lower-resolution Dame survey it continues to $\sim$14~pc and then flattens into a plateau, the behavior expected of a flow approaching a terminal velocity and not of rigid rotation (Sect.~\ref{sec:dame}).

The acceleration is consistent with the wind momentum flux from the Upper-Sco feedback flow coupled to the filament through cooling-driven entrainment at its wind-facing head. Although the formal survival criterion is comfortably satisfied ($t_\mathrm{cool,mix}/t_\mathrm{cc} \sim 10^{-3}$--$10^{-2}$), B44 lies in a colder, lower-velocity regime than the simulations of \citet{gronke_growth_2018, gronke_how_2020}, so we adopt their entrainment picture only qualitatively (Appendix~\ref{app:entrainment}). The implied kinematic timescale of the flow ($\sim$3~Myr) is well within the age of the Upper-Sco subgroup, supporting a dynamically maintained, quasi-steady flow rather than a single ballistic event.

We find further evidence that the wind-facing head (L1689) shows a bow shock: a deprojected flow Mach number of $\sim$2, a sonic-transition hierarchy from the bulk flow through the molecular envelope to the quiescent spine, and a density jump seen in the Herschel column density that matches the expected post-shock compression. More speculatively, the magnetic field may help to collimate and contain the entrained tail, and B44 may be streaming outward while rotating about its long axis. Sensitive dust-polarization mapping would test this possibility.

The acceleration we measure in B44 today roughly matches the rates measured along the Sco-Cen cluster chains \citep{posch_corona_2023, posch_physical_2025, miret-roig_tw_2025,grosschedl_evolution_2026}, which sample the same feedback over the preceding $\sim$15~Myr. B44 shows the same process still under way, in gas that has yet to form stars. The direction and magnitude of the flow are those predicted by the R-type filament scenario of \citet{alves_hp2_2025}, in which stellar feedback stretches dense gas radially outward from the feedback source into a head--tail morphology. More broadly, the technique should transfer. An inclination measured independently of the spectroscopy turns a line-of-sight velocity field into a physical one, so any coherent, well-resolved filament with a 3D distance constraint is now open to a direct measurement of gas acceleration in the star-forming ISM.

Two next steps would consolidate this picture. First, the lower-resolution Dame CO survey already indicates that the gradient levels off beyond $\sim$9~pc (Sect.~\ref{sec:dame}). Deeper, higher-resolution mapping across the full $\sim$20~pc extent of B44, beyond the $\sim$8~pc covered by the COMPLETE survey, would confirm this saturation and pin down its scale. Second, hydrodynamic simulations of cooling-driven entrainment matched to B44's parameters are needed to predict the detailed shape of $v(s)$ and the mass loading along the tail. A tailored simulation would turn the present demonstration into a quantitative test of the entrainment mechanism in the molecular-cloud regime.

\begin{acknowledgements}
We acknowledge fruitful discussions with Max Gronke that helped shape a possible interpretation for the formation of B44's tail.     
Co-funded by the European Union (ERC, ISM-FLOW, 101055318).
RW acknowledges support by the institutional project
RVO:67985815 and by the INTER-COST LUC24023 project of the
Czech Ministry of Education, Youth and Sports. The results in this paper were based on observations obtained with \textit{Herschel} and \textit{Gaia}, ESA science missions with instruments and contributions directly funded by ESA Member States, NASA, and Canada. This research has made use of NASA's Astrophysics Data System, the SIMBAD database and the VizieR catalog access tool operated at CDS, Strasbourg, France. This research used Astropy, a community-developed core Python package for Astronomy \citep{robitaille_astropy_2013}  and TOPCAT, an interactive graphical viewer and editor for tabular data \citep{taylor_topcat_2005}.
\end{acknowledgements}

\bibliographystyle{aa} 
\bibliography{bib}

\begin{appendix}

\section{PCA analysis}
\label{app:pca}

We derive the inclination angle $\alpha$ of B44 using a density-weighted PCA analysis of the 3D dust distribution as follows.

Let $n(\mathbf{p})$ denote the 3D-dust-derived volume density of hydrogen nuclei from \citet{leike_resolving_2020} \citep[using the conversion factor from][]{zucker_three-dimensional_2021}, and let $\{\mathbf{p}_i\}_{i=1}^{N}$ denote heliocentric Galactic Cartesian positions of the $N$ voxels in the vicinity of B44 that satisfy $n_i \equiv n(\mathbf{p}_i) > 30~\mathrm{cm}^{-3}$. Each voxel is assigned a weight equal to its local hydrogen nuclei volume density, $w_i \equiv n_i$.

We first compute the density-weighted centroid
\begin{equation}
    \mathbf{c} \;=\; \frac{\sum_{i} w_i\, \mathbf{p}_i}{\sum_{i} w_i},
    \label{eq:centroid}
\end{equation}
and the centered positions $\mathbf{r}_i \equiv \mathbf{p}_i - \mathbf{c}$. The $3\times3$ density-weighted covariance matrix is then
\begin{equation}
    \mathbf{C} \;=\; \frac{\sum_{i} w_i\, \mathbf{r}_i\, \mathbf{r}_i^{\top}}{\sum_{i} w_i},
    \qquad
    C_{jk} \;=\; \frac{\sum_{i} w_i\, r_{i,j}\, r_{i,k}}{\sum_{i} w_i},
    \label{eq:covariance}
\end{equation}

where $j, k \in \{x, y, z\}$. Each element $C_{jk}$ measures the density-weighted co-variation of the voxel positions along the $j$ and $k$ Cartesian axes; the diagonal entries $C_{xx}$, $C_{yy}$, $C_{zz}$ are the density-weighted variances of the voxel distribution along each axis. By construction $\mathbf{C}$ is real, symmetric, and positive semi-definite.

We then diagonalize $\mathbf{C}$ as
\begin{equation}
    \mathbf{C} \;=\; \mathbf{V}\,\mathbf{\Lambda}\,\mathbf{V}^{\top},
    \label{eq:eig}
\end{equation}

where $\mathbf{\Lambda} = \mathrm{diag}(\lambda_1, \lambda_2, \lambda_3)$ with $\lambda_1 \geq \lambda_2 \geq \lambda_3 \geq 0$, and the columns of $\mathbf{V}$ are the corresponding orthonormal eigenvectors $\{\hat{\mathbf{e}}_1, \hat{\mathbf{e}}_2, \hat{\mathbf{e}}_3\}$. Physically, each eigenvalue $\lambda_k$ is the density-weighted variance of the voxel positions projected onto $\hat{\mathbf{e}}_k$, so $\hat{\mathbf{e}}_1$ points along the direction of greatest spatial extent of the dense material. When the dust distribution is filamentary---as it is for B44---the largest eigenvalue strongly dominates ($\lambda_1 \gg \lambda_2, \lambda_3$), and $\hat{\mathbf{e}}_1$ is the natural mathematical representation of the filament's long axis. We therefore adopt $\hat{\mathbf{e}}_1$ as the 3D orientation of B44. For the mean dust cube, the $N = 707$ voxels above the density threshold give $(\lambda_1, \lambda_2, \lambda_3) = (84.9,\, 5.1,\, 2.2)$~pc$^2$, so that $\lambda_1 / \sum_k \lambda_k = 0.92$ and $\lambda_1/\lambda_2 = 16.5$. The dense material is thus strongly filamentary, with root-mean-square extents of $9.2$, $2.3$, and $1.5$~pc along the three principal directions, and is well represented by a single axis. The corresponding principal direction is $\hat{\mathbf{e}}_1 = (-0.445,\, -0.713,\, +0.542)$.

Let $\hat{\mathbf{r}}$ denote the unit vector along the line of sight (LOS) from the Sun to the filament,

\begin{equation}
    \hat{\mathbf{r}} \;=\; \bigl(\cos b_0 \cos l_0,\; \cos b_0 \sin l_0,\; \sin b_0\bigr),
    \label{eq:los}
\end{equation}

where $(l_0, b_0) = (354.7^{\circ},\, 15.1^{\circ})$ is the Galactic longitude and latitude of the LOS reference position adopted for B44. The plane of the sky at that position is, by definition, the plane perpendicular to $\hat{\mathbf{r}}$. The inclination angle $\alpha$ is the angle between the filament axis $\hat{\mathbf{e}}_1$ and this plane, i.e.,

\begin{equation}
    \alpha \;=\; \arcsin\bigl|\hat{\mathbf{e}}_1 \cdot \hat{\mathbf{r}}\bigr| \;=\; 90^{\circ} - \arccos\bigl|\hat{\mathbf{e}}_1 \cdot \hat{\mathbf{r}}\bigr|.
    \label{eq:alpha}
\end{equation}

The absolute value reflects the fact that the eigenvector $\hat{\mathbf{e}}_1$ has no intrinsic sign (both $\pm\hat{\mathbf{e}}_1$ are valid principal directions of $\mathbf{C}$); only the orientation of the axis matters, not the direction along it. We note that $\alpha$ depends only on the LOS direction $(l_0, b_0)$ and not on the distance $d$ to the reference point, since $\hat{\mathbf{r}}$ is a unit vector.

For the mean dust cube, $|\hat{\mathbf{e}}_1 \cdot \hat{\mathbf{r}}| = 0.223$ and thus $\alpha = 13^{\circ}$. Repeating the same PCA analysis on each of the 12 posterior realizations of the \citet{leike_resolving_2020} dust map yields a standard deviation of $2^{\circ}$ in $\alpha$, which we adopt as the uncertainty on the inclination.

\section{Error analysis}
\label{app:errors}

\begin{figure*}
    \centering
    \includegraphics[width=\linewidth]{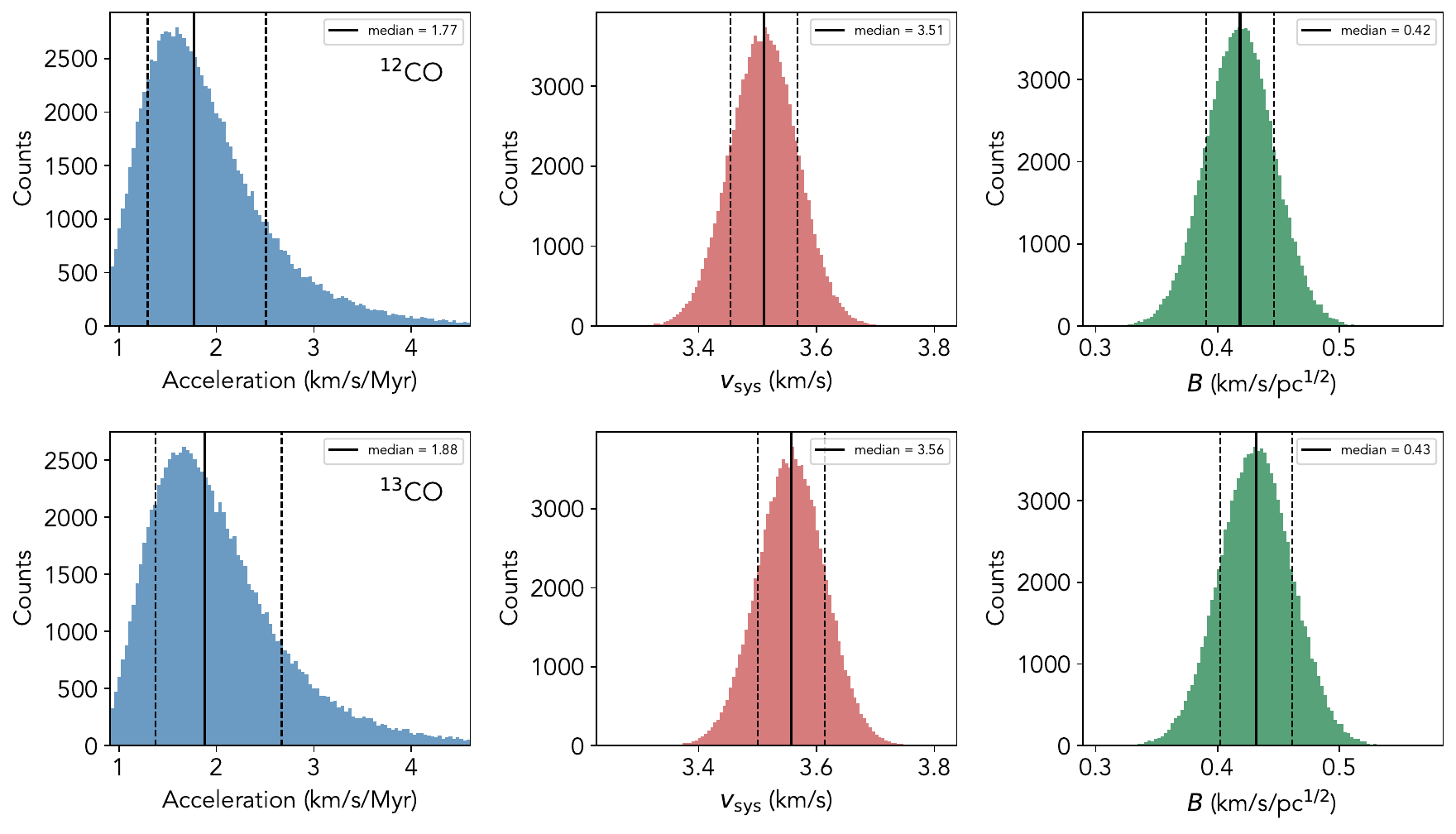}
\caption{Monte Carlo error analysis for the gas acceleration along B44, based on $10^5$ realizations sampling the covariance of the fit parameters ($v_\mathrm{sys}$, $B$) and the uncertainty on the inclination angle ($\alpha = 13^\circ \pm 2^\circ$). Top row: results for $^{12}$CO, with a median acceleration of $a = 1.77^{+0.74}_{-0.47}$~km/s/Myr, fitted systemic velocity $v_\mathrm{sys} = 3.51$~km/s, and velocity gradient parameter $B \approx 0.42$~km/s/pc$^{1/2}$. Bottom row: the same analysis applied to $^{13}$CO, yielding $a = 1.88^{+0.79}_{-0.51}$~km/s/Myr, $v_\mathrm{sys} = 3.56$~km/s, and $B \approx 0.43$~km/s/pc$^{1/2}$.
The acceleration distributions are right-skewed due to the nonlinear dependence $a = B^2 / (2\sin^2\alpha)$, so we report the median and 16th--84th percentile range rather than a symmetric uncertainty.}
\label{fig:mc_errors}
\end{figure*}

We estimated the uncertainty on the gas acceleration using a Monte Carlo approach with $10^5$ realizations. In each realization, we drew the fit parameters ($v_\mathrm{sys}$, $B$) from their joint posterior distribution, described by the covariance matrix returned by the least-squares fit, and independently sampled the inclination angle from a Gaussian distribution $\alpha = 13^\circ \pm 2^\circ$. For each draw, the acceleration was computed as $a = B^2/(2\sin^2\alpha)$. The resulting distributions are shown in Figure~\ref{fig:mc_errors}. Because $a$ depends nonlinearly on both $B$ (quadratically) and $\alpha$ (through $\sin^{-2}\alpha$), the acceleration distribution is right-skewed, and we report the median and 16th--84th percentile range rather than a symmetric uncertainty.

\section{Wind momentum budget and entrainment regime}
\label{app:momentum}

\subsection{Momentum flux}
\label{app:momentum_flux}

\begin{figure*}
    \centering
    \includegraphics[width=\linewidth]{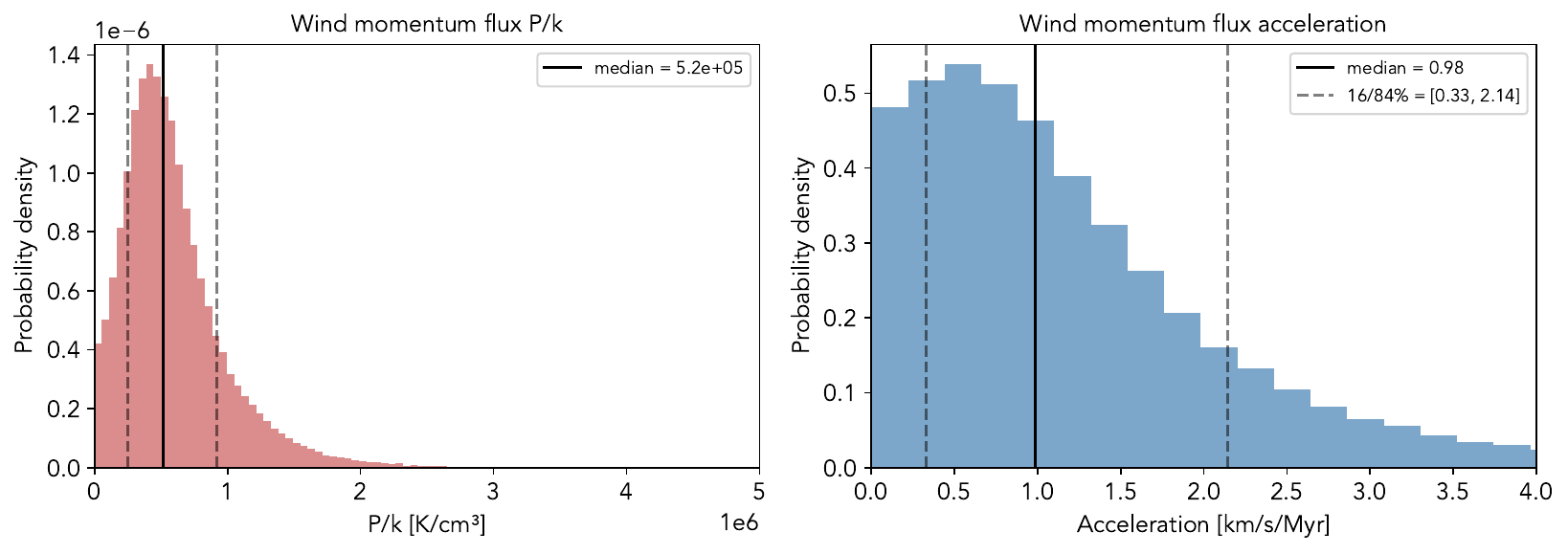}
    \caption{Monte Carlo analysis of the wind momentum flux from the Upper-Sco
feedback flow incident on the B44 filament, based on $10^5$ realizations sampling
the density of the impinging medium ($\rho \sim 10$~cm$^{-3}$), the
outflow velocity ($v \sim 20$~km/s), the filament cross-section
($A \sim 1$~pc$^2$), and the mass of the dense gas at the head ($M \sim 100$~M$_\odot$).
Left: the resulting distribution of momentum flux, with a median of
$P/k \approx 5.2 \times 10^5$~K\,cm$^{-3}$. Right: the corresponding
acceleration distribution, with a median of $a = 0.98^{+1.16}_{-0.66}$~km/s/Myr
(16th--84th percentile range). The distributions are right-skewed due to
the quadratic dependence of the momentum flux on velocity.
The median
acceleration agrees with the observed value ($\sim 1.8$~km/s/Myr, see
Sect.~\ref{sec:accel}) within the uncertainties, supporting the
interpretation that the Upper-Sco feedback flow can plausibly drive the
measured gas acceleration along B44.}
    \label{fig:momentum_mc}
\end{figure*}

As a consistency check, we estimate whether the momentum budget of the Upper-Sco feedback flow is sufficient to produce the observed acceleration. Adopting a fiducial density for the impinging medium of $\rho \sim 10$~cm$^{-3}$ and an outflow velocity of $v \sim 20$~km/s, the momentum flux is $\rho\,m_{\rm H}\,v^2 \approx 7 \times 10^{-12}$~Pa, corresponding to $P/k \approx 5 \times 10^5$~K\,cm$^{-3}$. The flow couples to the filament through entrainment 
and ablation at the wind-facing surface \citep[e.g.,][]{klein_hydrodynamic_1994, scannapieco_launching_2015, gronke_growth_2018}, with stripped material accelerated downstream along the filament axis.
Distributing the available momentum flux over an effective cross-section $A \sim 1$~pc$^2$ and the mass of the dense gas at the wind-facing head, $M \sim 100$~M$_\odot$ (rather than the full filament mass, since ablation concentrates the coupling at the head), yields a characteristic acceleration of $\sim 1$~km/s/Myr. A Monte Carlo treatment of the parameter uncertainties (Fig.~\ref{fig:momentum_mc}) gives a median of $0.98^{+1.16}_{-0.66}$~km/s/Myr, consistent with the observed value ($\sim 1.8$~km/s/Myr) within the uncertainties, which are dominated by the poorly constrained density of the impinging medium. A more definitive test requires independent constraints on the density and velocity of the impinging medium.

\subsection{Applicability of the cloud-entrainment model}
\label{app:entrainment}

The qualitative picture that motivates our interpretation, dense gas ablated at a wind-facing surface, mixed and partially recooled, then advected downstream into a tail, is the generic response of dense gas to an impinging flow. The cloud-crushing simulations with radiative cooling that have made this picture quantitative \citep{klein_hydrodynamic_1994, scannapieco_launching_2015, gronke_growth_2018, gronke_how_2020} were, however, computed in a regime far removed from B44, and we therefore use them only as a qualitative guide rather than as a calibrated model.

None of the conclusions we draw depends on the simulation-calibrated cooling physics. The wind carries more than enough momentum to drive the observed acceleration (Appendix~\ref{app:momentum_flux}). Ablation at a wind-facing surface followed by downstream advection is a generic outcome of a flow striking dense gas. And the observed velocity-profile shape, a quasi-flat head and an accelerating tail, follows from a residence-time argument, gas near the head having been exposed to the flow for less time than gas downstream, that is independent of the entrainment microphysics. We therefore adopt the entrainment picture qualitatively while treating the growth rates and steady-state classification of \citet{gronke_growth_2018, gronke_how_2020} as suggestive rather than established for B44. A dedicated simulation for the physical conditions of B44 is warranted.

\section{Caveats}
\label{app:caveats}

Two independent methods give a consistent acceleration of $a \approx 1.8$~km/s/Myr. A simple endpoint calculation using the deprojected velocity difference between L1689 and L1729 and the kinematic relation $v_\mathrm{flow}^2 = 2a\,\Delta s$ yields $a \approx 1.8$~km/s/Myr, matching the constant-acceleration fit over the full profile ($^{12}$CO: $\sim 1.8$~km/s/Myr; $^{13}$CO: $\sim 1.9$~km/s/Myr). We adopt the fit-based value as our primary result because it is constrained by all spatial bins rather than by the endpoints alone, and is therefore less sensitive to local perturbations at the head of the filament, where the direct interaction with the feedback flow may broaden and shift the velocity field.

To test the dependence of the fit on the length of the profile, we refit the model over the first 5~pc, 7~pc, and full (${\sim}8$~pc) of the deprojected velocity profile (Table~\ref{tab:subrange}). The fit is stable between 7~pc and the full length, with both tracers giving consistent accelerations ($^{12}$CO: $\sim 1.6$--$1.8$~km/s/Myr; $^{13}$CO: $\sim 1.9$~km/s/Myr).
When restricted to the inner 5~pc, however, the inferred acceleration drops to $\sim 0.4$--$0.6$~km/s/Myr because the velocity profile is approximately flat near the head before it begins to rise (see Figure~\ref{fig:PV12_fit}). The acceleration signal is therefore carried by the outer half of the observed profile, where gas velocities depart most strongly from the systemic value. In the entrainment picture (see Sect.~\ref{sec:feedback_flow}) this is the expected behavior: gas freshly stripped near the head sits close to the cloud rest frame, while gas at progressively larger downstream distances has spent longer in the flow and approached the wind velocity. The velocity profile is therefore better described as a quasi-flat head followed by an accelerating tail than as truly uniform acceleration from rest. We therefore report the whole-profile value ($a \approx 1.8$~km/s/Myr) as a characteristic average over the observed $\sim$8~pc, with the understanding that the instantaneous local acceleration varies along the filament.

\begin{table}[htbp]
\caption{Stability of the constant-acceleration fit with profile length. Parameters of $v = v_\mathrm{sys} + B\sqrt{\Delta s}$ fitted over the first $L_\mathrm{max}$ pc of the deprojected velocity profile. The fit is unstable below ${\sim}7$~pc because the velocity profile is nearly flat near the head; between 7~pc and the full length, both tracers give consistent accelerations.}
\label{tab:subrange}
\centering
\begin{tabular}{l c c c c}
\hline\hline
Tracer & $L_\mathrm{max}$ & $v_\mathrm{sys}$ & $B$ & $a$ \\
 & (pc) & (km/s) & (km/s/pc$^{1/2}$) & (km/s/Myr) \\
\hline
$^{12}$CO & 5.0 & $3.75 \pm 0.07$ & $0.23 \pm 0.04$ & $0.55 \pm 0.20$ \\
          & 7.0 & $3.54 \pm 0.06$ & $0.40 \pm 0.03$ & $1.59 \pm 0.25$ \\
          & 7.6 & $3.51 \pm 0.06$ & $0.42 \pm 0.03$ & $1.77 \pm 0.24$ \\
\hline
$^{13}$CO & 5.0 & $3.85 \pm 0.06$ & $0.19 \pm 0.04$ & $0.36 \pm 0.15$ \\
          & 7.0 & $3.55 \pm 0.06$ & $0.43 \pm 0.03$ & $1.90 \pm 0.26$ \\
          & 7.4 & $3.56 \pm 0.06$ & $0.43 \pm 0.03$ & $1.88 \pm 0.26$ \\
\hline
\end{tabular}
\end{table}

Additional caveats include: (i) the intensity-weighted mean velocity in each bin treats the emission as a single kinematic component, whereas the PV diagram (Figure~\ref{fig:PV13}) shows multiple components in some places, particularly near L1689; and (ii) protostellar outflows from the active star-forming region in L1689 may contaminate the velocity field near the head of the filament, although the gradient extending well beyond L1689 into quiescent regions argues that the large-scale flow is not driven by protostellar activity. The assumption of purely longitudinal flow and the alignment of B44 with the dominant Upper-Sco wind direction are discussed in Sect.~\ref{sec:alternatives}.

\section{Possible signatures of shocks and entrainment in B44's PV diagram}\label{sec:shocks}

\begin{figure*}[!tbp]
    \centering
    \includegraphics[width=\linewidth]{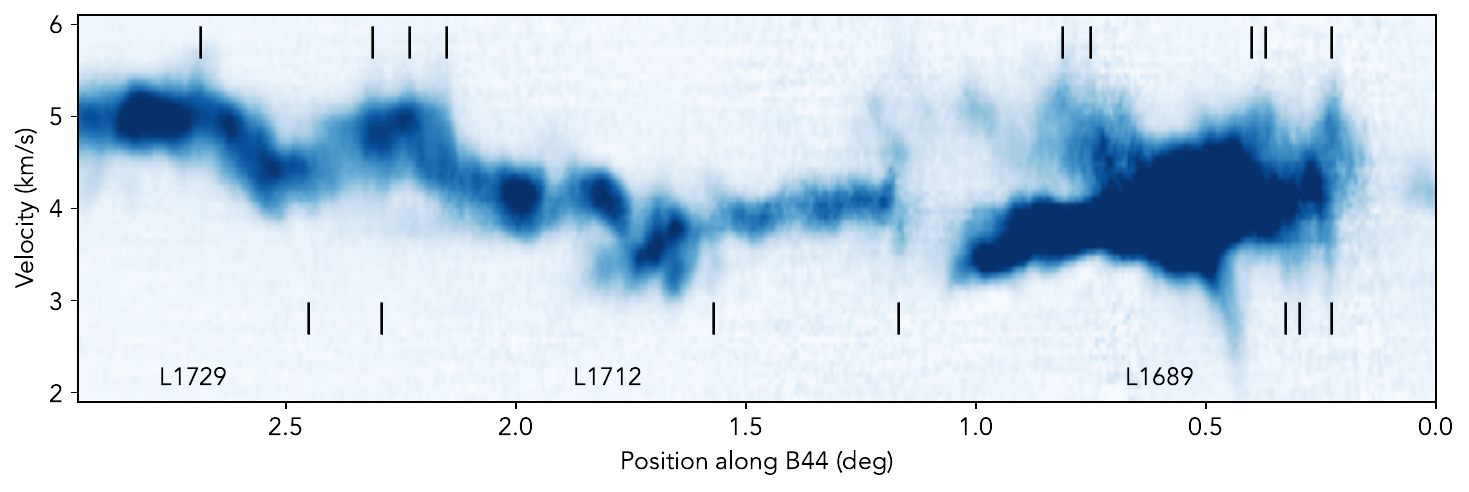}
    \caption{Same as Figure~\ref{fig:PV13}, but indicating a representative sample of narrow, nearly vertical features in the position-velocity diagram with black lines. These features indicate emission spanning a broad velocity range 
($\sim$1--2~km/s) at localized spatial positions along the filament. We speculate that these features are
consistent with shocks and gas entrainment features at the interfaces between 
the feedback flow from Upper-Sco and dense molecular clumps along B44.}
    \label{fig:pv_vert}
\end{figure*}

The position-velocity diagram of $^{13}$CO along B44 (Figure~\ref{fig:pv_vert}) reveals a number of narrow, nearly vertical features where emission spans a broad velocity range ($\sim$1--2~km/s) at essentially the same spatial position. These features are distinct from the overall monotonic velocity gradient along the filament and from the broader emission associated with individual clumps.

Vertical structure in a PV diagram indicates gas at a single projected location emitting over a wide range of velocities. In the context of the feedback-driven flow scenario, the most natural explanation is that these features trace shocks and gas entrainment at the interfaces between the low-density feedback flow from Upper-Sco and the dense molecular clumps along B44. Where the impinging flow encounters a clump surface, the resulting shock compresses gas into a spatially thin layer while producing emission across a broad velocity range, from the pre-shock ambient velocity through to the post-shock value. At the same interfaces, gas stripped from clump surfaces and entrained into the flow would produce intermediate-velocity emission bridging the clump velocity and the velocity of the surrounding medium. Both processes naturally generate the observed vertical PV morphology.

These features appear along the full extent of the filament, from L1689 to L1729, suggesting that the interaction between the feedback flow and the molecular gas is ongoing and not confined to the head of the filament. We note that near L1689, where active star formation is present, protostellar outflows may contribute to some of the observed velocity spread. However, the presence of similar features in the quiescent portions of the filament, far from any known protostellar activity, supports a feedback origin for most of the vertical structure in the PV diagram.

\section{Extended kinematics from the Dame CO survey}
\label{app:dame}
We followed the B44 flow past the COMPLETE coverage using the \citet{dame_milky_2001} DHT37 Ophiuchus $^{12}$CO cube ($\sim$0.25~deg, $\sim$0.6~pc per pixel). We extracted a position--velocity cut along the same filament spine with \texttt{pvextractor}, averaging over a 0.25~deg swath orthogonal to the path, and computed intensity-weighted mean velocities over 0--8~km/s following the same procedure and deprojection ($\alpha = 13^\circ$) as in Sect.~\ref{sec:velfield}, here with $3\sigma_v$ clipping.

The result is robust to the analysis choices. The profile is insensitive to the velocity window (0--6, 0--8, and 0--10~km/s give the same gradient and plateau). Wider averaging swaths blend in surrounding gas and mildly reduce the gradient amplitude, so the narrowest swath gives a lower limit on the acceleration that is consistent with the COMPLETE value. The plateau is not an artifact of beam dilution: a blend of the filament emission with lower-velocity ambient gas would imprint a second velocity component near the systemic $\approx 3.5$~km/s and pull the intensity-weighted centroid downward, but the spectra beyond $\sim$9~pc are single-peaked at $\approx 4.7$~km/s with no such component.

\end{appendix}
\end{document}